\documentclass[11pt]{article}
\usepackage{amssymb,latexsym,amsmath}

\usepackage{graphicx}
\usepackage{bm}

\begin{document}

\makeatletter

\@addtoreset{figure}{section}

\def\thefigure{\thesection.\@arabic\c@figure}

\def\fps@figure{h, t}

\@addtoreset{table}{bsection}

\def\thetable{\thesection.\@arabic\c@table}

\def\fps@table{h, t}

\def\theequation{\arabic{equation}}

\makeatother

\newtheorem{thm}{Theorem}[section]

\newtheorem{prop}[thm]{Proposition}

\newtheorem{lema}[thm]{Lemma}

\newtheorem{cor}[thm]{Corollary}

\newtheorem{defi}[thm]{Definition}

\newtheorem{rk}[thm]{Remark}

\newtheorem{exempl}{Example}[section]

\newcommand{\comment}[1]{\par\noindent{\raggedright\texttt{#1}

\par\marginpar{\textsc{Comment}}}}

\title{\textbf{\huge{A non incremental variational principle for brittle fracture}}}

\author{\Large{\textbf{G\'ery de Saxc\'e}}\\ \\
Univ. Lille, CNRS, Centrale Lille, UMR 9013 – LaMcube - \\
Laboratoire de m\'ecanique multiphysique multi\'echelle\\
F-59000, Lille, France, e-mail: gery.de-saxce@univ-lille.fr}


\maketitle                   




\begin{abstract}
The aim of the paper is to propose a paradigm shift for the variational approach of brittle fracture. Both dynamics and the limit case of statics are treated in a same framework. By contrast with the usual incremental approach, we use a space-time principle covering the whole loading and crack evolution. The emphasis is given on the modelling of the crack extension by the internal variable formalism and a dissipation potential as in plasticity, rather than Griffith's original approach based on the surface area.  The new formulation appears to be more fruitful for generalization than the standard theory.
\end{abstract}

\vspace{0.3cm}


\section{Introduction}

Many dynamical systems are subjected to energy loss resulting from dissipation, for instance collisions, surface friction, viscosity, plasticity, fracture and damage. On the other hand, deformations of solids and motions of fluids are modeled through constitutive laws. Due to collisions, brittle fracture and threshold effects, most dissipative laws are non smooth and multivalued. Moreover, experimental testing suggests that convexity is a keystone property of these phenomenological laws. Continuum mechanics with internal variables provides a convenient foundation to develop constitutive models that describe the inelastic behavior of various materials, despite the large differences in their physical nature and the relevant scale. The phenomenological law link the vector of internal variable rates  $\dot{\bm{\alpha}}$ to the dual force vector $\bm{A}$.
The theory of Generalized Standard Materials \cite{Halphen 1975} is based on an hypothesis of normal dissipation.

$$ \dot{\bm{\alpha}} \in \partial \varphi(\bm{A})
$$
where occurs the subdifferential of a convex and lower semicontinuous function $\varphi$, not everywhere differentiable. Equivalently, the law reads
$$ \bm{A} 
    = \mbox{argmax} \,(\left\langle \dot{\bm{\alpha}}, \bm{A} \right\rangle
      - \varphi (\bm{A}))
$$
or, introducing the Fenchel polar $\varphi^*$ \cite{Fenchel 1949}
$$ (\dot{\bm{\alpha}}, \bm{A})  
    = \mbox{argmin} \,(\varphi (\bm{A}) + \varphi^* (\dot{\bm{\alpha}})
    - \left\langle \dot{\bm{\alpha}}, \bm{A} \right\rangle)
$$
that does not favour one of the two dual variables. In the present work, the latest formulation is chosen as starting point, leading to variational methods and unconstrained or constrained optimization problems. Among them, Brezis-Ekeland-Nayroles principle \cite{Brezis Ekeland 1976} \cite{Nayroles 1976}, or in short BEN principle, is based on the time integration of the sum of dissipation potential $\phi$ and its Fenchel polar (analogous of Legendre polar for convex functions). Although used a few in the literature, this principle is noteworthy in the sense that it allows covering the whole evolution of the dissipative system.

In a previous paper \cite{Buliga 2009}, Buliga proposed the formalism of Hamiltonian inclusions, able to model dynamical systems with $1$-homogeneous dissipation potential (for laws such as brittle damage using Ambrosio-Tortorelli functional \cite{ambtor}). This formalism is a dynamical version of the quasi-static theory of rate-independent systems of Mielke (\cite{mielketh99}, \cite{mielke}, \cite{MR06b}). 

Latter on, Buliga and the author merged in a symplectic framework this formalism with BEN principle to extend it to dynamics \cite{SBEN}. The key-idea is to decompose additively the time rate $\dot{z}$ into reversible part $\dot{z}_R$ (the symplectic gradient) and dissipative or irreversible one $\dot{z}_I$, next to define the symplectic subdifferential $\partial^{\omega} \phi (z)$ of the dissipation potential. To release the restrictive hypothesis of $1$-homogeneity (in particular to address viscoplasticity), we introduce in this work the symplectic Fenchel polar $\phi^{*\omega}$, that allows to  build theoretical methods to model and analyse dynamical dissipative systems in a consistent geometrical framework with the numerical approaches not very far in the background. Numerical simulation with the BEN principle were performed for elastoplastic structures in statics \cite{Cao 2020, Cao 2021b} and in dynamics \cite{Cao 2021}.

Closer to this approach, we can cite the contributions of Aubin \cite{aubin2}, Aubin, Cellina and Nohel \cite{aubin}, Rockafellar \cite{rocka}, Stolz \cite{Stolz 1995}, which considered various extensions of Hamiltonian and Lagrangian mechanics. In the article \cite{bloch} by Bloch, Krishnaprasad, Marsden and Ratiu, Hamiltonian systems are explored with an added Rayleigh dissipation. A theory of quasistatic rate-independent systems is proposed by Mielke and Theil \cite{mielketh99}, Mielke \cite{mielke}, and developed towards applications in many papers, among them Mielke and Roub\'{\i}\v{c}ek \cite{MR06b}, see also Visintin \cite{visintin}. In (\cite{Grmela 1997}, \cite{Ottinger 1997}), Grmela and \"Ottinger proposed the framework GENERIC (General Equation for Non-Equilibrium Reversible-Irreversible Coupling), a systematic method to derive thermodynamically consistent evolution equations. A variational formulation of GENERIC is proposed in \cite{Manh Hong Duong 2013}. In \cite{Mielke 2011}, Mielke proposed a GENERIC formulation for Generalized Standard Materials quite similar to the one in the present paper. It would be worth to make the present formalism thermodynamic. In \cite{Stefanelli 2008}, Stefanelli  used Brezis-Ekeland-Nayroles variational principle to represent the quasistatic evolution of an elastoplastic material with hardening in order to prove the convergence of time and space-time discretizations as well as to provide some possible a posteriori error control. Finally, Ghoussoub and MacCann characterized the path of steepest descent of a non-convex potential as the global minimum of Brezis-Ekeland-Nayroles functional \cite{Ghoussoub 2004}.

Moreover, another advantage of Brezis-Ekeland-Nayroles principle is the easiness to be generalized. Indeed, it is worth to know that many realistic dissipative laws, called non-associated, cannot be cast in the mould of the standard ones deriving of a dissipation potential. To skirt this pitfall, the author proposed in \cite{saxfeng} a new theory based on a function called bipotential. It represents physically the dissipation and generalizes the sum of the dissipation potential and its Fenchel polar, reason for which extension of Brezis-Ekeland-Nayroles principle is natural. 
The applications of the bipotential approach to solid Mechanics are various: Coulomb's friction law \cite{sax CRAS 92}, non-associated Dr\"ucker-Prager  \cite{sax boussh IJMS 98} and Cam-Clay models \cite{Zouain 2010} in Soil Mechanics, cyclic Plasticity (\cite{sax CRAS 92},\cite{bodo sax EJM 01}, \cite{Bouby 2009}, \cite{Magnier 2006}, \cite{Bouby 2015}) and Viscoplasticity \cite{hjiaj bodo CRAS 00} of metals with non linear kinematical hardening rule, Lemaitre's damage law \cite{bodo},  the coaxial laws (\cite{dangsax},\cite{vall leri CONST 05}). Such kind of materials are called implicit standard materials. A synthetic review of these laws can be found in the two later references. It is also worth to notice that monotone laws but which does not admit a convex potential can be represented by Fitzpatrick's function \cite{Fitzpatrick 1988} which is a bipotential.

There is an abundant literature on the crack propagation criteria and our intention is not to draw here an exhaustive picture. For a deep survey, the reader is referred to \cite{Mroz 2010}. Let us cite only the most popular ones: 
the maximum tensile hoop stress criterion formulated in 1963 by Cherepanov \cite{Cherepanov 1963}, Erdogan and Sih \cite{Erdogan 1963},
the criterion of minimum strain energy density by Sih \cite{Sih 1973} in 1973, 
the principle of local symmetry formulated in 1974 by Goldstein and Salganik \cite{Goldstein 1974},
the maximum energy release rate criterion considered in 1974 by Hussain et al. \cite{Hussain 1974} and in 1978 by Wu \cite{Wu 1978},
and the generalized maximum energy release rate criterion \cite{He 1989, Gurtin 1998, Lebihain 2020, Lebihain 2021}.
These criteria give predictions more or less closed to the experimental results and the possible link with the variational approach is puzzling, at least apparently. This is one of the challenge we have to address. 

Our aim here is to study the propagation of already initiated cracks. Then we exclude of this paper the problem of the crack nucleation or initiation that would require extra information about stresses in the damaged region. Two models are used in the literature, the Coupled Criterion proposed par Leguillon \cite{Leguillon 2002} and the Cohesive Zone Model, originated in the pioneer works by Barenblatt \cite{Barenblatt 1959} and Dugdale \cite{Dugdale 1960}, next developed by Tvergaard and al \cite{Tvergaard 1992}, Xu \cite{Xu 1994} and, for the variational aspects, by Bourdin et al \cite{Bourdin 2008}. 

The paper is organized as follows. 
\begin{itemize}
    \item In Section 2, the initial crack and the loading history being known, the problem is to find the crack evolution. Instead of the usual modelling of the crack extension as a time-parameterized  family of surfaces, we propose to introduce a flow of a vector field on the final crack manifold.
    \item Section 3 is devoted to the definition of the variable of the problem  in the symplectic framework: the displacement and flow fields and the corresponding dynamic momenta.
    \item In Section 4, the symplectic BEN principle is applied to the fracture mechanics, that leads to the definition of a driving force, dual of the crack flow rate. 
    \item In Section 5, we propose a constitutive law for brittle fracture  in the form of a crack stability criterion and the normality law for the crack extension rule.
    \item We prove in Section 6 for structure with uniform toughness the equivalence with the classical variational approach in the form of a Munford-Shah functional.
    \item In Section 7, we deduce from the space-time principle an incremental version suitable for the usual step-by-step numerical approaches.
    \item In Section 8, we show how a suitable interpretation of the experimental data prove the relevancy of the normality law and the link with the principle of local symmetry.
    \item Section 9 is devoted to the calculation of the crack driving force  using the calculus of variation on the jet space of order one in the general case of dynamics.
\end{itemize}

\section{Modelling of the crack extension}

\subsection{Data}

\begin{figure}[h!]
	\centering
	\includegraphics[scale=1.30]{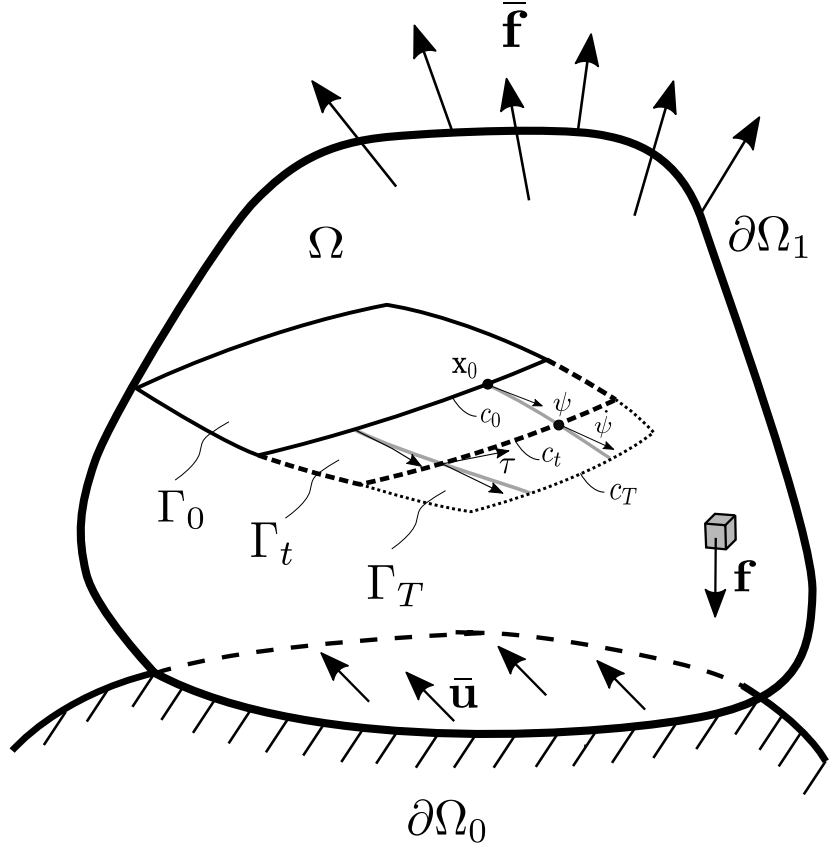}
	\caption{Crack Flow}
	\label{FigCrackFlow}
\end{figure}

Let $\displaystyle \Omega \subset \mathbb{R}^{n}$ be a bounded, open set, with piecewise smooth boundary $\partial \Omega$ occupied by the uncracked body (Figure \ref{FigCrackFlow}). Our aim is to study a crack extension during a time interval $\left[ 0, T \right] $

The data of the problem is:
\begin{itemize}
\item the initial crack $\Gamma_0$
\item the imposed displacements $\bar{\bm{u}}$ on the part  $\partial \Omega_0$ of the boundary (the supports)
\item the surface forces $\bar{\bm{f}}$ on the remaining part  $\partial \Omega_1$ 
\item the volume forces $\bm{f}$ 
\end{itemize}
All the external actions, $\bar{\bm{u}}, \bar{\bm{f}}$ and $ \bm{f}$, are prescribed during the time interval. No stress is transmitted through the crack. 

\subsection{Crack}
\label{SubSection Crack}
A crack bifurcates when its shape suffers a change of topology. The most common example is a crack  which develops in time new branches. During this phenomenon the number of crack fronts increases. In this paper we do not study the bifurcation of an existing crack. 

We denote $S$ the set of admissible surfaces, \textit{i.e.} closed countably $2$ rectifiable subsets of $\Omega$ without change of topology. The crack evolution is described by a time-parameterized family of surfaces (Figure \ref{FigCrackFlow})
$$ \Gamma : \left[ 0, T \right] \rightarrow S : t \mapsto \Gamma_t = \Gamma (t)
$$
such that $\Gamma(0) = \Gamma_0$ and the map $\Gamma$ is monotone increasing, $S$ being equipped with the inclusion order. The crack front $c_t$ at time $t$ is parameterized by the arc length $s$.

The cornerstone of the formulation is to model the crack extension by a flow on the cracked surface $\Gamma_T \backslash \Gamma_0$ during the time interval 
\begin{equation}
\left[ 0, T \right] \times c_0 \rightarrow \Gamma_T \backslash \Gamma_0 : (t, \bm{x}_0) \mapsto \bm{x} = \bm{\psi} (t, \bm{x}_0)
\label{crack flow}
\end{equation}
such that 
$$\bm{\psi} (\lbrace t \rbrace \times c_0)= c_t, \qquad
  \bm{\psi} (\left[ 0, t \right] \times c_0)= \Gamma_t \backslash \Gamma_0
$$
It is worth to remark that:
\begin{itemize}
    \item From a calculus viewpoint, the field $ \bm{\psi}$ represents both the crack and its evolution. The main advantage is that it is easier working with fields living in a functional space than with surfaces.
    \item In particular, our approach is well suited for numerical applications. $s \mapsto \bm{x}_0 = \bm{f} (s)$ being the parameterization of the initial crack $c_0$ by the arc length $s$, the cracked surface can be parameterized according to $(t,s) \mapsto \bm{x} = \tilde{\bm{\psi}} (t, s)  = \bm{\psi} (t, \bm{f} (s))$. The node $\bm{x}_{jk} = \tilde{\bm{\psi}} (t_j, s_k)$ being the position at time $t_j$ of the node $\bm{x}_{0k} = \bm{f} (s_k)\in c_0$ and $(s,t) \mapsto N_{jk} (t,s)$ the corresponding shape function, the cracked surface $\Gamma_T$ is parameterized by
    \begin{equation}
    \bm{x} = \tilde{\bm{\psi}} (t, s) =
        \sum_{j} \sum_{k}   N_{jk} (t,s) \, \bm{x}_{jk}
        \label{discretized crack flow}
    \end{equation}
    In the forthcoming variational formulation, the nodal values $\bm{x}_{jk} = \tilde{\bm{\psi}} (t_j, s_k)$ will be unknowns of the discretized problem. 
     \item A streamline $t \mapsto \bm{x} = \bm{\psi} (t, \bm{x}_0)$ does not correspond to the motion of a material particle initially at the position $\bm{x}_0 \in c_0$. The flow is not Lagrangian but Eulerian in the sense it represents just the evolution of the crack front. 
\end{itemize}

If the cracked surface is smooth enough, it is generated by a vector field $\dot{\bm{\psi}}$, solution on $\Gamma_T \backslash \Gamma_0$  of the ODE (Figure \ref{FigCrackFlow})
$$ \frac{d}{dt} (\bm{\psi} (t, \bm{x}_0)) 
   = \dot{\bm{\psi}}  (\bm{\psi} (t, \bm{x}_0))  \qquad 
    \mbox{with} \quad \bm{\psi} (0, \bm{x}_0) = \bm{x}_0
$$
As in the FEM method, it is easy to compute the vector field $\dot{\bm{\psi}}$ as the partial derivative of (\ref{discretized crack flow})  with respect to $t$. 

\section{Variables of the problem}

To do not enter into too complex formulations, we consider the case of an elastic body in small strains, although the generalization to the elastoplasticity and the finite strains is rather straightforward. The cracked solid at time $t$ is denoted $\Omega_t = \Omega \backslash \Gamma_t$. Within the body, the displacement field  at time $t$ is 
$$ \Omega_t  \rightarrow \mathbb{R}^3 : \bm{x} \mapsto \bm{u} (t, \bm{x})
$$

According to the symplectic BEN principle \cite{SBEN}, the variables of the problem of the elastodynamics are:
\begin{itemize}
    \item the field couple $\bm{\xi} = (\bm{u}, \bm{\psi}) $ of the displacement and the flow
    \item the momenta $\bm{\eta} = (\bm{p}, \bm{\pi}) $ where $\bm{p}$ is the classical linear momentum and $\bm{\pi}$ is the momentum associated to the flow
\end{itemize}
The system evolution is given by a curve 
$$ t \mapsto \bm{z} (t) = (\bm{\xi} (t), \bm{\eta} (t))
$$

\section{The symplectic BEN variational principle}

The duality between  the space $X$ of degrees of freedom $\bm{\xi}$ and the space $Y$ of momenta $\bm{\eta}$ has the form 
$$\langle \dot{\bm{\xi}}, \dot{\bm{\eta}} \rangle 
   = \langle \dot{\bm{u}}, \dot{\bm{p}} \rangle
     + \langle \dot{\bm{\psi}}, \dot{\bm{\pi}} \rangle
   = \int_{\Omega_t} \dot{\bm{u}}\cdot \dot{\bm{p}} \mbox{ d}^3 x 
     + \int_{c_t} \dot{\bm{\psi}}\cdot \dot{\bm{\pi}} \mbox{ d}s
$$ 
The symplectic form is the $2$-form field on $X \times Y$ defined by
$$ \omega (\dot{\bm{z}}, \dot{\bm{z}}') 
   = \omega ((\dot{\bm{\xi}},\dot{\bm{\eta}}) ,(\dot{\bm{\xi}}',\dot{\bm{\eta}}') ) 
   = \langle \dot{\bm{\xi}},\dot{\bm{\eta}}' \rangle 
     - \langle \dot{\bm{\xi}}',\dot{\bm{\eta}} \rangle 
$$
The total Hamiltonian of the structure is taken of the integral form
\begin{equation}
 H_t = H (t, \bm{z}) 
    = \int_{\Omega_t} \left\lbrace \dfrac{1}{2 \rho} \parallel \bm{p} \parallel^ 2 
                           + w(\nabla_s \bm{u}) - \bm{f} (t)\cdot\bm{u} \right\rbrace \mbox{ d}^3 x 
              - \int_{\partial \Omega_1} \bar{\bm{f}} (t)\cdot \bm{u} \mbox{ d}(\partial \Omega_1) 
\label{Hamiltonian}
\end{equation}
 The first term is the kinetic energy. The second one is  the elastic strain energy $w$ depending on the strain tensor $\bm{\varepsilon} = \nabla_s \bm{u} $ and the two latter terms are the works of the external forces. As usual, the stress tensor is
$$ \bm{\sigma} = \frac{\partial w}{\partial  \bm{\varepsilon}}
$$
A crack being a material discontinuity, we must distinguish  two material surfaces $\Gamma^+_t$ and $\Gamma^-_t$ that occupy the same position as $\Gamma_t$ but are the two sides of the crack and have opposite unit normal vectors, exterior to $\Omega_t$: $\bm{n}^+ = - \bm{n}^-$.

A curve is said admissible if it satisfies
\begin{itemize}
    \item the boundary conditions: 
    $$ \forall t \in \left[ 0, T \right], \quad  \bm{u} = \bar{\bm{u}}\;\;  \mbox{on} \;\; \partial \Omega_0, \quad 
    \bm{\sigma} \cdot \bm{n} = \bar{\bm{f}} \;\; \mbox{on} \;\; \partial \Omega_1, \quad 
    \bm{\sigma} \cdot \bm{n}^\pm = \bm{0} \;\; \mbox{on} \;\; \Gamma^\pm_t
    $$
    \item the initial conditions: $\bm{z} (0) = \bm{z}_0$
\end{itemize}

The SBEN formalism of dissipative media is based on the additive decomposition of the velocity into reversible and irreversible parts
$$ \dot{\bm{z}} = \dot{\bm{z}}_R + \dot{\bm{z}}_I
$$ 
As the variables of the problem are fields, we use functional derivatives
$$ \langle \delta\bm{\xi}, D_{\bm{\xi}} H  \rangle 
   + \langle D_{\bm{\eta}} H , \delta\bm{\eta} \rangle
    = \lim_{\epsilon \to 0} \frac{1}{\epsilon} 
      (H (t, \bm{z} + \epsilon \, \delta\bm{z}) 
      - H (t, \bm{z}) )
$$ 
The reversible part of the velocity is given by the symplectic gradient of the Hamiltonian (or Hamiltonian vector field)
$$ \dot{\bm{z}}_R = \nabla^\omega H 
   = (D_{\bm{\eta}} H, -  D_{\bm{\xi}} H)
   = ((D_{\bm{p}} H,  D_{\bm{\pi}} H), (- D_{\bm{u}} H, -  D_{\bm{\psi}} H))
$$
For the Hamiltonian (\ref{Hamiltonian}), we obtain
$$ D_{\bm{p}} H = \frac{\bm{p}}{\rho}, \qquad
   D_{\bm{\pi}} H = \bm{0}
$$
$$ - D_{\bm{u}} H = \nabla \cdot \bm{\sigma} + \bm{f}, \qquad
   - D_{\bm{\psi}} H = \bm{G}
$$
The last equation is just a definition of the \textbf{driving force} $\bm{G}$ of which the explicit expression will be discussed afterwards. 

In \cite{SBEN}, we formulated a general variational principle for the dynamical dissipative systems which claims that

\textbf{Symplectic BEN principle 1.} \textit{The natural evolution of the system minimizes the functional}
\begin{equation}
\Pi(\bm{z}) = \int_{0}^{T} \left\{ \phi(\dot{\bm{z}})      
    + \phi^{*\omega}(\dot{\bm{z}}_I) 
    - \omega (\dot{\bm{z}}_I, \dot{\bm{z}})  \right\} \mbox{ dt} 
\label{SBEN 1 Pi (z) =}
\end{equation}
\textit{among the admissible curves $t \mapsto \bm{z} (t) $ and the minimum is zero.} 

In this principle occurs the symplectic form $\omega$, the Hamiltonian through the irreversible part of the velocity $\dot{z}_I = \dot{z} - \nabla^\omega H $, a convex dissipation potential $\phi$ and its symplectic polar $\phi^{*\omega}$. We calculate now the detailed expression of this ingredients for the brittle fracture. We start with a simplifying hypothesis similar to the one introduced in \cite{SBEN} to recover the classical elastoplasticity. We claim that in the dissipation potential, the variables other than $\dot{\bm{\pi}} $ are ignorable
$$ \phi (\dot{\bm{z}}) = \varphi (\dot{\bm{\pi}})
$$
The symplectic polar is defined in \cite{SBEN} as
$$     \phi^{*\omega}(\dot{\bm{z}}') 
  = \sup \left[ \omega (\dot{\bm{z}}', \dot{\bm{z}})
       - \phi (\dot{\bm{z}}) \; : \; \dot{\bm{z}} \right]
$$
Under the previous hypothesis, one has
$$ \phi^{*\omega}(\dot{\bm{z}}') =
 \sup \left[ 
   \left\langle \dot{\bm{u}}', \dot{\bm{p}} \right\rangle
 + \left\langle \dot{\bm{\psi}}', \dot{\bm{\pi}} \right\rangle
 - \left\langle \dot{\bm{u}}, \dot{\bm{p}}' \right\rangle
 - \left\langle \dot{\bm{\psi}}, \dot{\bm{\pi}}' \right\rangle
 - \varphi (\dot{\bm{\pi}}) \; : \; \dot{\bm{z}} \right]
$$
that gives
$$ \phi^{*\omega}(\dot{\bm{z}}') =
   \chi_{\left\lbrace \bm{0} \right\rbrace} (\dot{\bm{u}}' )
   + \varphi^* (\dot{\bm{\psi}}' )
   + \chi_{\left\lbrace \bm{0} \right\rbrace} ( \dot{\bm{p}}')
   + \chi_{\left\lbrace \bm{0} \right\rbrace} ( \dot{\bm{\pi}}')
$$
where $\chi_K$ is the indicatory function of the set $K$, equal to $0$ in $K$ and $+\infty$ otherwise, and $\varphi^*$ is the Fenchel polar of $\varphi$. The second term in the functional (\ref{SBEN 1 Pi (z) =}) becomes:
$$ \phi^{*\omega}(\dot{\bm{z}}_I) =
   \chi_{\left\lbrace \bm{0} \right\rbrace} \left(\dot{\bm{u}} - \frac{\bm{p}}{\rho} \right) 
   + \varphi^* (\dot{\bm{\psi}} )
   + \chi_{\left\lbrace \bm{0} \right\rbrace} ( \dot{\bm{p}} - \nabla \cdot \bm{\sigma} - \bm{f})
   + \chi_{\left\lbrace \bm{0} \right\rbrace} ( \dot{\bm{\pi}} - \bm{G})
$$
As the minimum of the functional is the finite value zero, it will be reached when the arguments of the indicatory functions vanish. Then we may take the zero value of the indicatory functions in the functional  while we introduce extra corresponding constraints
\begin{equation}
    \bm{p} = \rho \, \dot{\bm{u}}, \qquad
 \nabla \cdot \bm{\sigma} +  \bm{f} = \dot{\bm{p}}, \qquad
 \dot{\bm{\pi}} = \bm{G}
 \label{extra constraints}
\end{equation}
Next we can transform the last term in the functional, taking into account the velocity decomposition, the linearity and the antisymmetry of the symplectic form
\begin{equation}
    - \omega (\dot{\bm{z}}_I, \dot{\bm{z}})
 = - \omega (\dot{\bm{z}} - \dot{\bm{z}}_R, \dot{\bm{z}})
 = - \omega (\dot{\bm{z}}, \dot{\bm{z}})
   + \omega (\dot{\bm{z}}_R, \dot{\bm{z}})
 = \omega (\dot{\bm{z}}_R, \dot{\bm{z}})
 \label{- omega (dot(z)_I, dot(z)) =}
\end{equation}
or in detail
$$ - \omega (\dot{\bm{z}}_I, \dot{\bm{z}})
= \left\langle \frac{\bm{p}}{\rho}, \dot{\bm{p}} \right\rangle
 - \left\langle \dot{\bm{u}}, \nabla \cdot \bm{\sigma} +  \bm{f}  \right\rangle
 - \left\langle \dot{\bm{\psi}}, \bm{G} \right\rangle
$$
Owing to the constraints (\ref{extra constraints}), it holds
$$ - \omega (\dot{\bm{z}}_I, \dot{\bm{z}})
= \left\langle \dot{\bm{u}} , \dot{\bm{p}} - \nabla \cdot \bm{\sigma} -  \bm{f}  \right\rangle
 - \left\langle \dot{\bm{\psi}}, \bm{G} \right\rangle
= - \left\langle \dot{\bm{\psi}}, \bm{G} \right\rangle
$$
Moreover, thanks to the former and latter constraints in (\ref{extra constraints}), the momenta can be eliminated from the functional and the intermediate constraint becomes
$$ \nabla \cdot \bm{\sigma} +  \bm{f} = \rho \, \ddot{\bm{u}} 
$$
while the initial condition on the linear momentum can be transformed into an initial condition on the velocity because $\bm{p} (0) = \rho \, \dot{\bm{u}} (0)$. Taking into account these transformations, the minimum can be searched only on the space of the degrees of freedom $\bm{u}$ and $ \bm{\psi}$.
Then we obtain a second version of the variational principle

\textbf{Symplectic BEN principle 2.} \textit{The natural evolution of the system minimizes}
\begin{equation}
\Pi(\bm{\xi}) = \int_{0}^{T} \left\{ 
\varphi (\bm{G}) + \varphi^* (\dot{\bm{\psi}})
- \left\langle \dot{\bm{\psi}}, \bm{G} \right\rangle
\right\} \mbox{ d}  t
\label{SBEN 2 Pi (x) =}
\end{equation}
\textit{among the admissible curves $t \mapsto \bm{\xi} (t) $ such that} 
$ \nabla \cdot \bm{\sigma} +  \bm{f} = \rho \, \ddot{\bm{u}} $
\textit{and the minimum is zero.} 

It could seem puzzling that the displacement does not appear explicitly in the expression of the functional but in fact $\bm{u}$  and $\bm{\psi}$ are coupled in the minimization problem:
\begin{itemize}
    \item $\bm{\psi}$ is controlled by $\bm{G}$ which depends on $\bm{u}$ as derivative of the Hamiltonian. 
    \item $\bm{u}$ must satisfy the constraint and the admissibility conditions that are defined on $\Omega_t = \Omega \backslash \Gamma_t$ controlled by  $\bm{\psi}$.
\end{itemize}

\section{Constitutive laws for brittle fracture}

As discussed in the Introduction, most of the crack stability criteria in the literature, in particular the most popular, are of local nature and not variational. Nevertheless, Strifors proposed in \cite{Strifors 1973} to predict the onset of brittle fracture and the extension direction thanks to the crack extension force that can be identified to the driving force $ \bm{G}$. Now we show how to cast it into the mold of our variational approach. Based on the works by Strifors  \cite{Strifors 1973} and Hellen et al. \cite{Hellen 1975}, we propose a simple constitutive law. At least for plane problems, it seems reasonable to think that the influence on the fracture extension of the projection of $\bm{G}$ onto the tangent to the crack front can be neglected. Then we introduce the deviatoric force
$$ \bm{G}_\perp = \bm{G} - (\bm{G} \cdot \bm{\tau}) \, \bm{\tau}
$$
where $\bm{\tau}$ is the unit tangent vector to the crack front in the direction of increasing arc length (Figure \ref{FigCrackFlow}). As in the FEM method, it is easy to compute $\bm{\tau}$ from the partial derivative of (\ref{discretized crack flow}) with respect to the arc length $s$ of $c_0$. If we assume that the toughness of the material is \textbf{isotropic}, the critical energy release rate $G_c$ being a parameter measuring it, we introduce the convex \textbf{crack stability domain}
$$ K = \left\lbrace \bm{G} \;\; \mbox{such that} \;\; 
     \parallel \bm{G}_\perp \parallel \leq G_c
       \right\rbrace
$$
and the \textbf{crack extension rule} 
$$ \dot{\bm{\psi}} \in \partial \varphi (\bm{G}) 
     = \partial \chi_K (\bm{G})
$$
that is the \textbf{normality law}
\vspace{0.5cm}

\begin{tabular}{|lll|}
\cline{1-3}
 & & \\
\textbf{if} &  $\bm{G} \in \mathring{K}$ \textbf{then} &  \\
 & $ \dot{\bm{\psi}} = \bm{0} $ & ! crack stability \\
 \textbf{else} &  $\bm{G} \in \partial K$ \textbf{and} 
                    $ \exists \lambda \geq 0 , \qquad 
    \dot{\bm{\psi}} = \lambda \, \bm{G}_\perp   $     
 & ! crack extension \\
 & & \\
 \cline{1-3}
\end{tabular}

\vspace{0.5cm}
and $\dot{\bm{\psi}} $ is perpendicular to $\bm{\tau}$ when the crack extends. The Fenchel polar is the support function
$$ \varphi^* (\dot{\bm{\psi}}) 
   = \int_{c_t} G_c \parallel \dot{\bm{\psi}} \parallel \, \mbox{d}s
$$
where $s$ is the arc length of $c_t$. With this choice of constitutive law, we particularize the variational principle in the form

\textbf{Symplectic BEN principle 3.} \textit{The natural evolution of the system minimizes}
\begin{equation}
\Pi(\bm{\xi}) = \int_{0}^{T} \left\{ 
   \int_{c_t} G_c \parallel \dot{\bm{\psi}} \parallel \, \mbox{d}s
- \left\langle \dot{\bm{\psi}}, \bm{G} \right\rangle
\right\} \mbox{ d} t
\label{SBEN 3 Pi (x) =}
\end{equation}
\textit{among the admissible curves $t \mapsto \bm{\xi} (t) $ such that} 
$ \nabla \cdot \bm{\sigma} +  \bm{f} = \rho \, \ddot{\bm{u}} $, $ \parallel \bm{G}_\perp \parallel \leq G_c$
\textit{and the minimum is zero.} 

\section{Link with the classical variational approach to fracture}

First we recall that the last term in (\ref{SBEN 3 Pi (x) =}) is in fact the last one of the integrand of (\ref{SBEN 1 Pi (z) =}). Owing to (\ref{- omega (dot(z)_I, dot(z)) =}), we have
$$     - \omega (\dot{\bm{z}}_I, \dot{\bm{z}})
    = \omega (\dot{\bm{z}}_R, \dot{\bm{z}})
    = \langle D_{\bm{\eta}} H, \dot{\bm{\eta}} \rangle 
   - \langle \dot{\bm{\xi}}, - D_{\bm{\xi}} H   \rangle
    = \dot{H} - D_t H
$$
Besides the crack surface element comprised between the crack fronts $c_t$ and $c_{t + dt}$, and the streamlines of $\bm{x}_0$ and $\bm{x}_0 + d\bm{x}_0$ is an infinitesimal parallelogram with two sides supported by the vectors $ d_1 \bm{x} = \bm{\tau} \, ds$ and $d_2 \bm{x} = \dot{\bm{\psi}}\, dt$. Its area is
$$ d\Gamma = \parallel \bm{\tau} \times \dot{\bm{\psi}}\parallel \, ds \, dt 
=  \parallel \bm{\tau}\parallel \,
  \parallel \dot{\bm{\psi}}\parallel \,
  \mid \sin{(\bm{\tau}, \dot{\bm{\psi}})} \mid
  \, ds \, dt 
   = \parallel \dot{\bm{\psi}}\parallel
  \, ds \, dt 
$$
because $\bm{\tau}$ is a unit vector and $\dot{\bm{\psi}} $ is perpendicular to $\bm{\tau}$.

If we assume that the toughness properties of the material are \textbf{homogeneous}, the critical energy release rate is uniform 
$$ G_c (\bm{x}) = G_c
$$
The area of a surface will be denoted by $\mid \bullet \mid $.
The variational principle can be recast as

\textbf{Symplectic BEN principle 4.} \textit{The natural evolution of the system minimizes}
\begin{equation}
\Pi(\Gamma, \bm{u}) = G_c \, \mid \Gamma_T \backslash \Gamma_0  \mid
    -  \int_{0}^{T} D_t H \, \mbox{ d}  t
     + H_T - H_0
\label{SBEN 4 Pi (x) =}
\end{equation}
\textit{among the admissible curves $t \mapsto (\Gamma (t), \bm{u} (t)) $ such that} 
$ \nabla \cdot \bm{\sigma} +  \bm{f} = \rho \, \ddot{\bm{u}} $, $ \parallel \bm{G}_\perp \parallel \leq G_c$
\textit{and the minimum is zero.} 

Two cases must be considered in the applications:
\begin{itemize}
    \item If the loading is controlled by the forces, the  structural evolution is in general dynamic and the functional to minimize is
    $$ \Pi(\Gamma, \bm{u}) = G_c \, \mid \Gamma_T \backslash \Gamma_0  \mid
    -  \int_{0}^{T} D_t H \, \mbox{ d}  t
     + H_T - H_0
    $$
    \item If the loading is slow and controlled by the displacements (for instance, $\bar{\bm{f}}$ is null, the support is divided into two disjoint parts, one of them is fixed to the foundation, the other one is loaded), the evolution is quasi-static and the functional to minimize is reduced to 
    $$ \Pi (\Gamma, \bm{u})= G_c \, \mid \Gamma_T \backslash \Gamma_0  \mid
     + H_T - H_0
    $$
    Regardless of the constant $H_0$, we recover the Munford-Shah functional \cite{Munford 1989, Francfort 1998, Buliga 1999, Bourdin 2008}
\end{itemize}

\section{Incremental method}
 
Our goal now is to find an incremental or step-by-step formulation as a by-product of our general principle. As usual, the time interval $\left[ 0, T \right]$ is divided into $N$ small sub-intervals  $\left[ t_n , t_{n+1} \right]$ with $t_0 = 0$ and $t_N = T$. the value of a variable $\bm{a}$ at time $t_n$ is denoted $\bm{a}_n$. The time step is $\Delta t = t_{n+1} - t_n$ and $\Delta \bm{a} = \bm{a}_{n+1} - \bm{a}_n$ is the increment of $\bm{a}$. 

$\bm{\xi}_n = (\bm{u}_n, \bm{\psi}_n)$ being known from the initial conditions for $n=0$ or from the previous increment otherwise and the time step being prescribed, the problem is to calculate 
 $\bm{\xi}_{n+1} $ or equivalently $\Delta \bm{\xi}$. The idea is to apply the 
 symplectic BEN principle 3 to the sub- interval 
 $\left[ t_n , t_{n+1} \right]$. The time step being small, the functional can be approximated by considering the integrand is constant. As for the radial return algorithm in plasticity, we use the implicit scheme by evaluating the crack extension rate $\dot{\bm{\psi}}$ and the driving force $\bm{G}$ at the end of the step. The crack extension increment is
 \begin{equation}
 \Delta \bm{\psi} = \dot{\bm{\psi}}_{n+1} \, \Delta t
     = \lambda \, \bm{G}_{n+1} \Delta t
     \label{implicit scheme}
 \end{equation}

 Hence the incremental principle reads
 
 \textbf{Symplectic BEN principle $3^*$.} \textit{The natural evolution of the system minimizes}
\begin{equation}
\Delta \Pi(\bm{\xi}_{n+1}) = 
   \int_{c_{t_{n+1}}}  G_c \parallel \Delta \bm{\psi} \parallel \, \mbox{d}s
- \left\langle \Delta \bm{\psi}, \bm{G}_{n+1} \right\rangle
\label{SBEN 3 Delta Pi (x) =}
\end{equation}
\textit{among the} $ \bm{\xi}_{n+1} $ \textit{such that} 
$ \nabla \cdot \bm{\sigma}_{n+1} +  \bm{f}_{n+1} = \rho \, \ddot{\bm{u}}_{n+1} $, $ \parallel (\bm{G}_{n+1})_\perp  \parallel \leq G_c$
\textit{and the minimum is zero.} 

Likewise, the last version of the principle in the previous section gives rise to the incremental form

 \textbf{Symplectic BEN principle $4^*$.} \textit{The natural evolution of the system minimizes}
\begin{equation}
\Delta \Pi(\Gamma_{n+1}, \bm{u}_{n+1}) = 
   G_c \mid \Delta \Gamma \mid
 -  (D_t H)_{n+1} \, \Delta  t
     + \Delta H
\label{SBEN 4 Delta Pi (x) =}
\end{equation}
\textit{among the couples} $(\Gamma_{n+1}, \bm{u}_{n+1})$ \textit{such that} 
$ \nabla \cdot \bm{\sigma}_{n+1} +  \bm{f}_{n+1} = \rho \, \ddot{\bm{u}}_{n+1} $, $ \parallel (\bm{G}_{n+1})_\perp  \parallel \leq G_c$
\textit{and the minimum is zero.} 

If the loading is slow and controlled by the displacements, the evolution is quasi-static and the functional to minimize is reduced to
$$ \Delta \Pi(\Gamma_{n+1}, \bm{u}_{n+1}) = 
   G_c \mid \Delta \Gamma \mid
     + \Delta H
$$
and we recover Buliga's incremental formulation \cite{Buliga 1999}.
 
 \section{Comparison of the constitutive law with experiments}

\begin{figure}[h!]
	\centering
	\includegraphics[scale=0.60]{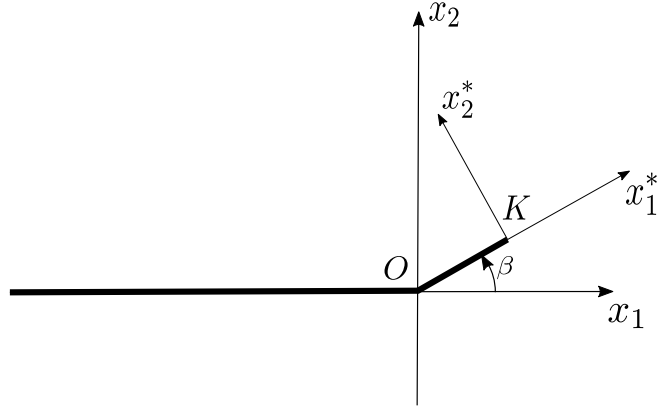}
	\caption{Kinked crack}
	\label{FigKink}
\end{figure}

 The aim of this section is to check the validity of the normality law with respect to the experimental results on PMMA specimens \cite{Richard 1984}. A good test is to considered an initial straight crack in a plate which, under mixed mode loading, may extend suddenly in a direction deviating of an angle $\beta$ from the original one by kinking (Figure \ref{FigKink}). The length of the kink crack is taken very small. The idea is to use the incremental formulation of the previous section with only one step. To simplify the notations, the initial value $a_0$ of a given quantity is denoted simply $a$ while the value $a_1$ at the end of the step is denoted $a^*$. In Hellen et al. \cite{Hellen 1975}, Euler explicit scheme is used that can read with the simplified notations
 $$    \Delta \bm{\psi} = \dot{\bm{\psi}} \, \Delta t 
       = \lambda \, \bm{G} \Delta t
 $$
 In the frame of origin $O$ at the initial crack tip, the axis $Ox_1$ in the direction ahead the initial crack and the axis $Ox_2$ perpendicular to $Ox_1$ within the plane (Figure \ref{FigKink}), the driven force is given in terms of stress intensity factors (SIFs) by
 $$G_1 =\frac{1}{\bar{E}} \, (K^2_{I} + K^2_{II}) ,\qquad
   G_2 =-  \frac{1}{\bar{E}} \, 2\, K_{I} K_{II}
 $$
 where $\kappa = 3 - 4 \, \nu$ in plane strain, $\kappa = (3 - \nu) / (1 + \nu)$ in plane stress and 
 $$ \frac{1}{\bar{E}} = \frac{(1 + \nu) \, (1 + \kappa)}{4 \, E}
 $$
The deviation angle $\beta$ of the kink crack is given by the slope of the driven force
 $$ \beta = \tan^{-1} \left(- \frac{2\, K_{I} K_{II}}{K^2_{I} + K^2_{II}}\right)
 $$
 In mode $I$, the formula gives $\beta = 0^\circ$, according to the experience. For $K_{I} = K_{II}$, $\beta = - 45^\circ$ but the experimental values are above $- 49^\circ$. In mode $II$, $\beta = 0^\circ$ while the experimental values are in the range from $ - 70^\circ$ to $- 74^\circ$. Clearly, the predictions are disastrous. Our opinion is that the constitutive law must not be \textit{a priori} rejected but it is the explicit scheme which is problematic.
 
 The problem of the kinked crack were studied by many authors. It is an awkward  problem of Elasticity and only approximated formula were proposed to express the SIFs at the kink crack tip in terms of the SIFs at the initial crack tip. At the limit of vanishing length of the kink crack, we adopt in the sequel the expression proposed in \cite{Cotterell 1980}
 $$ K^*_{I} (\beta) = \cos^3 \frac{\beta}{2} \, K_{I} 
              - 3 \, \sin \frac{\beta}{2} \, \cos^2 \frac{\beta}{2} \, K_{II}
 $$
 $$ K^*_{II} (\beta) =  \sin \frac{\beta}{2} \, \cos^2 \frac{\beta}{2} \, K_{I} 
               + \cos \frac{\beta}{2} \, \left(1 - 3 \,  \sin^2 \frac{\beta}{2} \right) \, K_{II}
 $$
 However, it is noticeable that a correction is proposed in \cite{Fett 2002} for the largest values of the $\mid \beta \mid$ and the dependency with respect to the length of the kink crack is taken into account in \cite{Leblond 1989, Amestoy 1992} but these improvements will be not considered here. With the simplified notations, the implicit scheme (\ref{implicit scheme}) reads
 \begin{equation}
    \Delta \bm{\psi} = \dot{\bm{\psi}}^* \, \Delta t 
       = \lambda \, \bm{G}^* \Delta t
     \label{implicit scheme star}
 \end{equation}
  In the frame of origin $K$ at the kink crack tip, the axis $Kx^*_1$ in the direction ahead the kink crack and the axis $Kx^*_2$ perpendicular to $Kx^*_1$ within the plane (Figure \ref{FigKink}), the driven force is given in terms of stress intensity factors (SIFs) by
 $$G^*_1 =\frac{1}{\bar{E}} \, \left[(K^*_{I})^2 + (K^*_{II})^2\right] ,\qquad
   G^*_2 = -  \frac{1}{\bar{E}} \, 2\, K^*_{I} K^*_{II}
 $$
 The normality law (\ref{implicit scheme star}) entails 
 \begin{equation}
  \bar{E}\, G^*_2 = - 2\, K^*_{I} K^*_{II} = 0, \qquad
 \bar{E}\, \parallel \bm{G}^* \parallel 
   =(K^*_{I})^2 + (K^*_{II})^2 ,\qquad
     \label{implicit scheme star 2}
 \end{equation}
 To satisfy the former condition, two scenarios may be considered for the crack extension:
 \begin{enumerate}
     \item \textbf{Scenario 1.} The kink crack inclination is $\beta_1$, solution of $ K^*_{II} (\beta_1) =  0 $, and $\bar{E}\, \parallel \bm{G}^* \parallel =  (K^*_{I} (\beta_1))^2  $
     \item \textbf{Scenario 2.} The kink crack inclination is $\beta_2$, solution of $ K^*_{I} (\beta_2) =  0 $, and $\bar{E}\, \parallel \bm{G}^* \parallel =  (K^*_{II} (\beta_2))^2  $     
 \end{enumerate}
 According to the stability criterion, the crack extends for the inclination $\beta_i$ with the maximum driven force magnitude. Then  scenario 1 is realized if $(K^*_{I} (\beta_1))^2 > (K^*_{II} (\beta_2))^2  $  and  scenario 2 otherwise.
 
 The admissible solutions must be searched in the interval $\left] - \pi, \pi \right[$. The mode $I$ singularity disappearing in compression, $K_{I}$ is positive. In mode $II$, the SIF $K_{II}$ may be positive or negative, depending on the sign of the shear loading. Before tackling the mixed mode, let us consider the limit case:
 \begin{itemize}
     \item \textbf{Mode I.} For scenario 1, there is a unique admissible solution $\beta_1 = 0^\circ$ and $(K^*_{I} (\beta_1))^2 = K^2_{I}$. For scenario 2, there is no admissible solution. Then the scenario 1 occurs, according to the experimental observations.
\end{itemize}
 Let us examine now the general case ($K_{I} \, K_{II} \neq 0 $). For scenario 1, there are two admissible solutions
 $$ \beta_{1\pm} = \sin^{-1} 
    \left(\frac{\frac{K_{II}}{K_{I}} \pm 3 \, \frac{K_{II}}{K_{I}} \, 
      \sqrt{8 \, \left( \frac{K_{II}}{K_{I}}\right)^2 + 1}}
     {9 \, \left( \frac{K_{II}}{K_{I}}\right)^2 + 1}\right) 
 $$
 For scenario 2, there is a unique admissible solution 
 $$ \beta_2 = 2 \, \tan^{-1} \left( \frac{K_{I}}{3 \, K_{II}}\right)
 $$
 In particular,  let us discuss the case:
 \begin{itemize}
     \item \textbf{Mixed mode} $\bm{K}_{\bm{I}} = \bm{K}_{\bm{II}} > 0$. For scenario 1, there are two admissible solutions: 
      $\beta_{1-} = \sin^{-1} (- 4/5) = - 53.13^\circ$ then $(K^*_{I} (\beta_{1-}))^2 = (1.847 \, K_{I})^2$ and $\beta_{1+} = \sin^{-1} (1) =  90^\circ$ then $(K^*_{I} (\beta_{1+}))^2 = (- 1.14 \, K_{I})^2$.  For scenario 2, there is  a unique admissible solution $\beta_2 = 36.87^\circ$ and $(K^*_{II} (\beta_2))^2 = (0.948 \, K_{I})^2$. Comparing the values of $(K^*_{I} (\beta_{1-}))^2$, $(K^*_{I} (\beta_{1+}))^2$ and $(K^*_{II} (\beta_2))^2 $, we conclude that the scenario 1 occurs with the angle $\beta = \beta_{1-} = - 53.13^\circ$ closed to the experimental values ($- 49^\circ$ for $K_{II} = 0.874 \, K_{I}$ and values in the range from $- 56^\circ$ to $- 59^\circ$ for $K_{II} = 1.142 \, K_{I}$).
 \end{itemize}
 
 In this respect, it is worth to remark that, if the sign of $K_{II}$ is reversed, the sign of the kink angles are reversed too : $\beta_{1-} = 53.13^\circ$,  $\beta_{1+} =  - 90^\circ$ and $\beta_2 = - 36.87^\circ$ but the corresponding values of $(K^*_{I} (\beta_{1-}))^2$, $(K^*_{I} (\beta_{1+}))^2$ and $(K^*_{II} (\beta_2))^2 $ are the same. We conclude that the sign of the kink angle is reversed: $\beta = 53.13^\circ$. 
 
 Finally, let us deal with the other limit case:
 \begin{itemize}
\item \textbf{Mode II.} For scenario 1, there are two admissible solutions 
      $\beta_{1 \pm} = 2\, \sin^{-1} (\pm \, 1/ \sqrt{3}) = \pm \,  70.47^\circ$ and $(K^*_{I} (\beta_{1 \pm}))^2 = 4 \, K^2_{II} / 3$. For scenario 2, there is  a unique admissible solution $\beta_2 = 0^\circ$ and $(K^*_{II} (\beta_2))^2 = K^2_{I}$. Then the scenario 1 occurs and the sign of the kink angle depends of the one of mode $II$.  For $K_{II}> 0$, the prediction $\beta = - 70.47^\circ$ is just above the experimental values.
 \end{itemize}
 
 \textbf{Comment 1.} On the basis of the experimental data, Richard proposed an empirical criterion in terms of the SIFs at the original crack tip
 $$ \frac{K_{I}}{K_{Ic}} + \left( \frac{K_{II}}{K_{IIc}} \right)^2 = 1
 $$
 According to (\ref{implicit scheme star 2}), the crack extends when
 $$ (K^*_{I})^2 + (K^*_{II})^2 = \bar{E}\, G_c
 $$
 that gives, for the scenario 1 of mode I, $K^2_{I} = \bar{E}\, G_c $ then
 $$ K^2_{Ic} = \bar{E}\, G_c
 $$
 and, for the scenario 1 of mode II,  $4 \, K^2_{II} / 3 = \bar{E}\, G_c = K^2_{Ic}$ then
 $$ \frac{K_{IIc}}{K_{Ic}} = \frac{\sqrt{3}}{2} = 0.86
 $$
 a value better than $\sqrt{2/3} = 0.81$ proposed in \cite{Fett 2002}, by comparison with the experimental data covering the range from $0.88$ to $0.95$. Moreover, our prediction for the mixed mode $K_{I} = K_{II}$ leads to values $\frac{K_{I}}{K_{Ic}} = 0.54 $ and $\frac{K_{II}}{K_{IIc}} = 0.62$ closed to the curve of Richard criterion.

 \textbf{Comment 2.} At least in this example, the relevant scenario corresponds always to the \textbf{Principle of Local Symmetry} $ K^*_{II} (\beta) =  0 $ proposed in 1974 by Goldstein and Salganik \cite{Goldstein 1974}. This principle is valid at least for a kinked straight crack with vanishing kink crack but there can be no assurance that it is valid for arbitrary cracks and we recommend to replace it by the  normality law proposed in this paper. 
 
 \textbf{Comment 3.} The present criterion should not be confused with the \textbf{Maximum Energy Release Rate criterion}, proposed in 1974 by Hussain et al. \cite{Hussain 1974}, for which the maximum must be searched among all the directions of the crack extension while with our criterion the maximum is found only among the directions satisfying the normality law, according to the different scenarios. As argued in \cite{Chambolle 2009}, the questions of when and how a crack propagates should be simultaneously investigated and the energy conservation is not sufficient for such a task.

 \textbf{Comment 4.} This example shows also that the implicit scheme must be preferred to the explicit scheme in fracture mechanics.

 \section{Calculation of the crack driving force}
 
 As we are working  at time $t$ in this Section, the dependency with respect to the time will be omitted. In order to obtain the expression of $\bm{G} $, we use a special form of the calculus of variation performed on the jet space of order one (\cite{Aldaya}, \cite{Edelen}, \cite{Mangiarotti}). For more details about the jet spaces, the reader is referred for instance to \cite{Saunders}. The first jet prolongation of the smooth function $\bm{u} : \Omega_t \rightarrow \mathbb{R}^3 $ is the function $j^1 \bm{u}$ from $\Omega_t$ into the jet bundle $J^1 ( \Omega_t, \mathbb{R}^3 ) $ such that 
$$ j^1 \bm{u}\ (\bm{x}) = \left( \bm{x}, \bm{u}\ (\bm{x}), \nabla \bm{u} (\bm{x}) \right) 
$$
The Hamiltonian (\ref{Hamiltonian}) at time $t$ has the form
$$ H 
    = \int_{\Omega_t} h (\bm{x}, \bm{u}, \nabla \bm{u}, \bm{p}) \mbox{ d}^3 x 
              - \int_{\partial \Omega_1} \bar{\bm{f}} 
                 \cdot \bm{u} \mbox{ d}(\partial \Omega_1) 
$$
$$ H 
    = \int_{\Omega_t} h (j^1 \bm{u}, \bm{p}) \mbox{ d}^3 x 
              - \int_{\partial \Omega_1} \bar{\bm{f}} \cdot \bm{u} \mbox{ d}(\partial \Omega_1) 
$$

\begin{figure}[h!]
	\centering
	\includegraphics[scale=0.80]{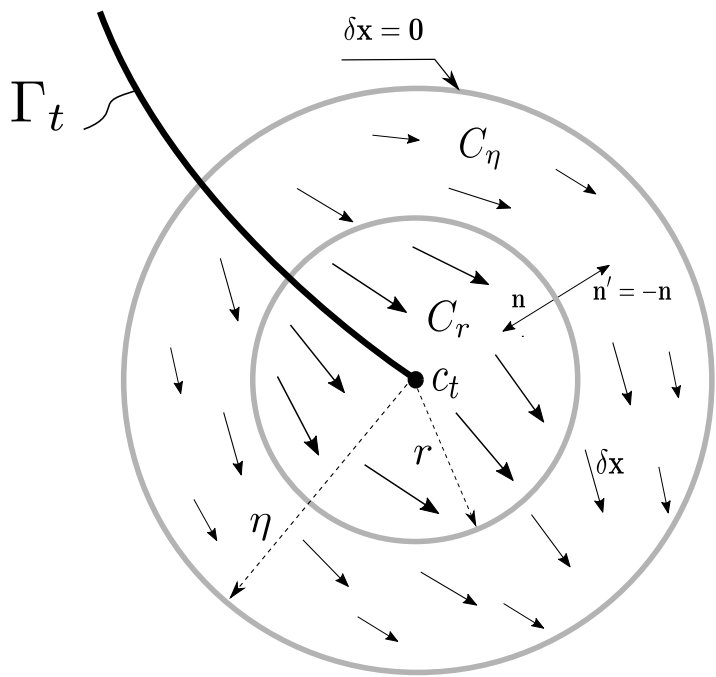}
	\caption{Cylindrical neighbourhood}
	\label{FigCylindricalNeighbourhood}
\end{figure}

The new viewpoint which consists in replacing the original field $\bm{u}$ by its first jet prolongation $j^1 \bm{u}$ leads to perform variations not only on the field and its derivatives but also on the variable $\bm{x}$. 
We want to calculate the functional derivative of $H$ in the direction $\delta \bm{\psi}$ defined on the crack front $c_t$. The idea is to prolong it locally by a field $\delta \bm{x}$ on an open cylindrical neighbourhood $C_\eta$ of radius $\eta$ around the crack front, the field vanishing at radius $\eta$ and beyond (Figure \ref{FigCylindricalNeighbourhood}).  To avoid the singularity of the stress field, we replace $C_\eta$ by  $C_{\eta, r} = C_\eta \backslash \bar{C}_r $ where $C_r$ is an open cylindrical neighbourhood  of radius $r < \eta$ around the crack front. Next, we pass to the limit $\eta \rightarrow r \rightarrow 0 $.

We consider now a new parameterization given by a regular map $\bm{x} = \bm{\theta} (\bm{y})$ of class $C^1$ on $C'_{\eta, r}= \bm{\theta}^{-1} (C_{\eta, r} ) $ and we perform the variation of the function $\bm{\theta}$, the new variable being $\bm{y}$. The variation of the action is
\begin{equation}
   \delta H 
    = \delta \int_{C'_{\eta, r}} h (\bm{\theta} (\bm{y}), \bm{u}, \nabla_{\bm{y}} \bm{u} \cdot \nabla_{\bm{x}} \bm{y}, (\nabla_{\bm{x}} \bm{y})^T \cdot \bm{p}') \, \det (\nabla_{\bm{y}} \bm{x}) \mbox{ d}^3 y   
\label{delta H = 1}
\end{equation}
where $ \bm{p}'$ is the linear momentum in $\bm{y}$ coordinates. We do not give here the details of the calculations that can be found in Appendix A but only indicate the sketch of the method and the result. As usual in calculus of variation, we integrate by part. Next we consider the limit case where the function $\bm{\theta}$ approaches the identity of $\Omega_t$, $\bm{y}$ approaches $\bm{x}$, $\bm{p}'$ approaches $\bm{p} $ and $C'_{\eta, r}$ approaches $C_{\eta, r}$.
Finally, the variation of the Hamiltonian has the form
\begin{equation}
 \delta H 
    = \int_{\partial C_{\eta, r}} \bm{n}\cdot \bm{T} \cdot \delta \bm{x} \mbox{ d}(\partial C_{\eta, \rho}) 
    - \int_{C_{\eta, r}} \left( \nabla \cdot \bm{T} + \bm{f}\right) \cdot \delta \bm{x} \mbox{ d}^3 x
    \label{delta H = 3}
\end{equation}
where 
\begin{equation}
 \bm{T} = h \, \bm{1}_{\mathbb{R}^3} - \bm{\sigma} \cdot \nabla \bm{u} - \dot{\bm{u}} \otimes \bm{p}
 \label{def T}
\end{equation}
The contribution on the cylindrical part $\partial C_{\eta, r}$ of radius $\eta$ vanishes as $\delta \bm{x}$. When $\eta$ approaches $r$, the contributions on the remaining parts of $\partial C_{\eta, r}$ excepted the cylindrical part of radius $r$, vanish too. Let $\bm{n}'$ the outward unit vector  to $\partial C_r$, opposite to the outward unit vector $\bm{n}$ to $\partial C_{\eta, r}$ (Figure \ref{FigCylindricalNeighbourhood})
$$ \bm{n}' = - \bm{n}
$$
Then using the polar coordinates $r, \theta$, one has
$$ \delta H 
     = - \lim_{r \rightarrow 0} \int_{\partial C_{r} \cap \partial C_{\eta, r}} \bm{n}' \cdot \bm{T} \cdot \delta \bm{x}\;  r \mbox{ d}\theta \mbox{ d} s
$$ 
Finally, recalling that the functional derivative is a density on the crack front, the value of the driving force corresponding to the arc length $s$ is 
\begin{equation}
\bm{G}  = - D_{\delta \bm{x}} H
     = \lim_{r \rightarrow 0} \int_{c_{r, s}} \bm{n}' \cdot \bm{T} \;  r  \mbox{ d}\theta 
     \label{G = limit}
\end{equation}
where $c_{r, s}$ is the line at the intersection between the cylindrical-shaped surface $\partial C_{r} \cap \partial C_{\eta, r}$ and the plane normal to the crack front $c_t$ at the arc length $s$. This limit is called by Gurtin et al. \cite{Gurtin 1998} a tip integral and denoted
$$ \bm{G} 
     =  \oint_{\mbox{\tiny{tip}}} \bm{n}' \cdot \bm{T} 
$$
but we do not recover the last term of (\ref{def T}) in  Gurtin's  and Stolz's \cite{Stolz 1995} expressions  of  $\bm{T}$. In this form, the driving force appears as a generalization of Rice-Eshelby $J$ integral. In statics, this quantity has the well known property of path-independence and it is relevant to wonder whether this property remains true in dynamics. The elastic stress field being singular at the crack front, the volume force field may be neglected when $r$ approaches zero. In Appendix B, it is shown that for the natural evolution of the system
\begin{equation}
\nabla \cdot \bm{T} 
    = (\nabla \bm{u}) \cdot \dot{\bm{p}}
      - (\nabla \cdot \dot{\bm{u}}) \, \bm{p}
      - 2 \, (\nabla_s \bm{p})  \cdot \dot{\bm{u}}
    \label{nabla cdot T + f =}
\end{equation}
Unfortunately, this expression does not vanish in general but only in the limit case of statics where we recover the classical Rice-Eshelby identity
$$ \nabla \cdot \bm{T} 
    = \nabla \cdot \left[ w  \, \bm{1}_{\mathbb{R}^3} - \bm{\sigma} \cdot \nabla \bm{u} \right]   = \bm{0}
$$
Then except the particular case of statics, there is no path-independence of the integral in dynamics and passing to the limit is required in (\ref{G = limit}).

 \section{Conclusions and perspectives}
 
 In this work, we presented a particular version of the non incremental symplectic BEN principle. It can be generalized in various ways:

\begin{itemize}
    \item The Hamiltonian $H$ depends on $\bm{u}, \bm{p}$ and $\bm{\psi}$ (through $\Omega_t = \Omega \backslash \Gamma_t$) but not on $\bm{\pi}$. It was a bias motivated by the wish to recover classical formulations of the fracture mechanics but we could give up this restrictive assumption by introducing an explicit dependence of the Hamiltonian on  $\bm{\pi}$.
    \item For the same motivation, the dissipation potential $\phi$ depends on $\dot{\bm{\pi}}$ but not on $\dot{\bm{u}}, \dot{\bm{p}} $ and $\dot{\bm{\psi}} $. Nevertheless there is nothing preventing us from lifting this restriction, in particular concerning  $\dot{\bm{\psi}} $.
    \item The potential $\varphi$ could fully depend on $\bm{G}$, not only on $\bm{G}_\perp$. It could be also generalized to anisotropic behaviours with symmetry group techniques.  
    \item For easiness, we considered only 1-homogeneous potential $\varphi^*$ in this work to illustrate the method but we can add extra terms to represent for instance the dependence of the critical energy release rate with respect to the crack extension vector as in \cite{Chopin 2018, Kolvin 2015, Lebihain 2020, Lebihain 2021}. The crack extension by fatigue can be also modelled thanks to homogeneous potentials of degree linked to the crack growth law (\cite{Lemaitre Chaboche EN}, \cite{Bourdin 2008}).
    \item We could easily take into account more sources of dissipation such as damping, plasticity, damage and so on, by considering in the symplectic formalism new internal variables and the corresponding momenta, and by simply adding terms in the dissipation potential $\phi$.
    \item Sometimes there is no getting away from experimental facts and the kinetic law is non associated. For such atypical events, the present principle can be generalized, replacing the sum of the dissipation potential and its Fenchel polar by a bipotential. 
\end{itemize}

We are convinced of the interest of this non incremental principle in computational structural mechanics because the error can be controlled uniformly on the whole evolution, in contrast to incremental methods which accumulate the errors and degrade the accuracy over time.  

We hope that our application of the normality law to the kinked crack allows to take forward the discussion in the literature on the choice of the crack stability criterion. 

In the future, we plan to develop numerical methods based on the present approach. We already have given a hint about the discretization of the crack flow at the end of Section \ref{SubSection Crack}. In order to avoid a cumbersome remeshing, an efficient technique is the XFEM where the standard displacement-based approximation is enriched by incorporating both discontinuous fields and the near tip asymptotic fields (\cite{Moes 1999}, \cite{Dolbow 2000a}, \cite{Dolbow 2000b}, \cite{Moes 2002}). 

\vspace{0.5cm}
\textbf{\Large{Acknowledgements}}

\vspace{0.3cm}

The author would like to thank Marius Buliga, Djimedo Kondo, Abdelbacet Oueslati, Pierre Gosselet, Céline Bouby and Long Cheng for their questions and comments during seminars that allowed to improve the paper.


\vspace{0.5cm}

\textbf{\Large{Appendix A}}

\vspace{0.3cm}

To calculate the expression of the crack driven force, we start from the variation of the Hamiltonian (\ref{delta H = 1})  
$$    \delta H 
    = \delta \int_{C'_{\eta, r}} h (\bm{\theta} (\bm{y}), \bm{u}, \nabla_{\bm{y}} \bm{u} \cdot \nabla_{\bm{x}} \bm{y}, (\nabla_{\bm{x}} \bm{y})^T \cdot \bm{p}') \, \det (\nabla_{\bm{y}} \bm{x}) \mbox{ d}^3 y  
$$
and we take the variation with respect to the fields $\bm{u}$ and $\bm{\theta}$
$$   \delta H 
    =  \int_{C'_{\eta, r}} 
    \lbrace
    \left[
    Tr \left(\bm{\sigma}\,  \delta (\nabla_{\bm{y}} \bm{u} \cdot \nabla_{\bm{x}} \bm{y} )\right)
    + \dot{\bm{u}} \cdot \delta ((\nabla_{\bm{x}} \bm{y})^T \cdot \bm{p}')
    - \bm{f} \cdot \delta \bm{x}
    \right] \, \det (\nabla_{\bm{y}} \bm{x})
$$
\begin{equation}
 \qquad\qquad\qquad\qquad\qquad\qquad\qquad\qquad\qquad\qquad\qquad\qquad
    + h \, \delta (\det (\nabla_{\bm{y}} \bm{x}))
    \rbrace \,\mbox{ d}^3 y   
    \label{delta H = 2}
\end{equation}
Next we calculate the variation of the field derivative in terms of the derivative of the variation of the field
$$ \delta (\nabla_{\bm{y}} \bm{u} \cdot \nabla_{\bm{x}} \bm{y} ) 
   = - \nabla_{\bm{y}} \bm{u} \cdot \nabla_{\bm{x}} \bm{y}\cdot 
      \nabla_{\bm{y}} (\delta \bm{x}) \cdot \nabla_{\bm{x}} \bm{y}
$$
$$ \delta (\det (\nabla_{\bm{y}} \bm{x})) 
    = Tr \left( \nabla_{\bm{y}} (\delta \bm{x}) 
       \cdot \mbox{adj} (\nabla_{\bm{y}} \bm{x})\right)
$$
where $adj (\bm{A})$ is the transposed of the matrix of cofactors of $\bm{A}$
$$ \mbox{adj} (\bm{A}) = \det (\bm{A}) \, \bm{A}^{-1}
$$ 
Likewise, one has
$$ \delta ((\nabla_{\bm{x}} \bm{y})^T \cdot \bm{p}') 
  = - (\nabla_{\bm{x}} \bm{y}\cdot 
      \nabla_{\bm{y}} (\delta \bm{x}) \cdot \nabla_{\bm{x}} \bm{y})^T \cdot \bm{p}' 
   = - (\nabla_{\bm{x}} \bm{y})^T \cdot 
      (\nabla_{\bm{y}} (\delta \bm{x}))^T \cdot (\nabla_{\bm{x}} \bm{y})^T \cdot \bm{p}' 
$$
Inserting the three previous expression into (\ref{delta H = 2}) and after simple manipulations, we obtain
$$   \delta H 
    =  \int_{C'_{\eta, r}} 
    \lbrace
    Tr \left( \mbox{adj} (\nabla_{\bm{y}} \bm{x}) \cdot
       \left( h\, \bm{1}_{\mathbb{R}^3}
       -  \bm{\sigma} \cdot \nabla_{\bm{y}} \bm{u} \cdot \nabla_{\bm{x}} \bm{y} - (\dot{\bm{u}} \otimes \bm{p}') \cdot \nabla_{\bm{x}} \bm{y} \right) \cdot
       \nabla_{\bm{y}} (\delta \bm{x})
    \right) 
$$
$$ \qquad\qquad\qquad\qquad\qquad\qquad\qquad\qquad\qquad\qquad\qquad\qquad
    - \det (\nabla_{\bm{y}} \bm{x}) \, \bm{f} \cdot \delta \bm{x}
    \rbrace \,\mbox{ d}^3 y   
$$
Integrating by parts leads to
$$   \delta H 
    = \int_{\partial C'_{\eta, r}} 
     \bm{n} \cdot \mbox{adj} (\nabla_{\bm{y}} \bm{x}) \cdot
       \left( h\, \bm{1}_{\mathbb{R}^3}
       -  \bm{\sigma} \cdot \nabla_{\bm{y}} \bm{u} \cdot \nabla_{\bm{x}} \bm{y} - (\dot{\bm{u}} \otimes \bm{p}') \cdot \nabla_{\bm{x}} \bm{y} \right) \cdot
       \delta \bm{x} \, \mbox{d} (\partial C'_{\eta, r})
$$
$$   -  \int_{C'_{\eta, r}} 
    \lbrace
    \nabla \cdot \left( \mbox{adj} (\nabla_{\bm{y}} \bm{x}) \cdot
       \left( h\, \bm{1}_{\mathbb{R}^3}
       -  \bm{\sigma} \cdot \nabla_{\bm{y}} \bm{u} \cdot \nabla_{\bm{x}} \bm{y} - (\dot{\bm{u}} \otimes \bm{p}') \cdot \nabla_{\bm{x}} \bm{y} \right) 
       \right) 
$$
$$ \qquad\qquad\qquad\qquad\qquad\qquad\qquad\qquad\qquad\qquad\qquad\qquad
    + \det (\nabla_{\bm{y}} \bm{x}) \, \bm{f} \rbrace \cdot \delta \bm{x}
     \,\mbox{ d}^3 y   
$$
Considering the limit case where the function $\bm{\theta}$ approaches the identity of $C_{\eta, r}$, $\bm{y}$ approaches $\bm{x}$, $\bm{p}'$ approaches $\bm{p} $ and $C'_{\eta, r}$ approaches $C_{\eta, r}$, we obtain (\ref{delta H = 3})
$$ \delta H 
    = \int_{\partial C_{\eta, r}} \bm{n}\cdot \bm{T} \cdot \delta \bm{x} \mbox{ d}(\partial C_{\eta, \rho}) 
    - \int_{C_{\eta, r}} \left( \nabla \cdot \bm{T} + \bm{f}\right) \cdot \delta \bm{x} \mbox{ d}^3 x
$$
where occurs the  tensor (\ref{def T}) 
$$ \bm{T} = h \, \bm{1}_{\mathbb{R}^3} - \bm{\sigma} \cdot \nabla \bm{u} - \dot{\bm{u}} \otimes \bm{p}
$$ 
\vspace{0.5cm}

\textbf{\Large{Appendix B}}

\vspace{0.3cm}

Because of the stress singularity, the volume force that are regular may be consider as uniform near the crack front. In one hand, the variation of the Hamitonian density $h$ in the direction $\delta \bm{x}$ is
\begin{equation}
   \delta h = \nabla h \cdot \delta \bm{x}
            = \nabla \cdot ( h \, \bm{1}_{\mathbb{R}^3} )
              \cdot \delta \bm{x}
    \label{delta h = 1}
\end{equation}
In the other hand, according to (\ref{Hamiltonian}), $h$ is a function of $\bm{x}$ through $j^1 \bm{u}$ and $\bm{p}$
$$ h =  \dfrac{1}{2 \rho} \parallel \bm{p} \parallel^ 2 
                           + w(\nabla_s \bm{u}) - \bm{f} (t)\cdot\bm{u} 
$$
Using the chain rule and assuming the volume force are uniform, one has
$$ \delta h = Tr (\bm{\sigma} \cdot \nabla (\delta \bm{u})) 
             + \dot{\bm{u}} \cdot \delta \bm{p}
             -  \bm{f} \cdot \bm{u}
$$
Using the linear momentum balance (intermediate constraints in (\ref{extra constraints}) which is satisfied for the natural evolution of the system), the first term of the right hand member becomes
$$ Tr (\bm{\sigma} \cdot \nabla (\delta \bm{u})) 
    = \nabla \cdot (\bm{\sigma} \cdot \delta \bm{u}) 
      - (\nabla \cdot \bm{\sigma}) \cdot \delta \bm{u}
    = \nabla \cdot (\bm{\sigma} \cdot \delta \bm{u})
      + (\bm{f} - \dot{\bm{p}}) \cdot \delta \bm{u}
$$
Hence
$$ \delta h = \nabla \cdot (\bm{\sigma} \cdot \delta \bm{u})
       - \dot{\bm{p}} \cdot \delta \bm{u}
       + \dot{\bm{u}} \cdot \delta \bm{p}
$$
The infinitesimal variations of the displacement and linear momentum fields resulting from an infinitesimal arbitrary variation $\delta \bm{x}$ being
$$ \delta \bm{u} = \nabla \bm{u} \cdot \delta \bm{x}, \qquad
   \delta \bm{p} = \nabla \bm{p} \cdot \delta \bm{x}
$$
and in particular if the field $\delta \bm{x}$ is uniform
\begin{equation}
 \delta h = \left[ \nabla \cdot (\bm{\sigma} \cdot \nabla \bm{u})
       - \dot{\bm{p}} \cdot \nabla \bm{u}
       + \dot{\bm{u}} \cdot \nabla \bm{p} 
       \right] \cdot \delta \bm{x}
    \label{delta h = 2}
\end{equation}
The expressions  (\ref{delta h = 1}) and (\ref{delta h = 2})  being equal for every $\delta \bm{x}$, we obtain
$$ \nabla \cdot ( h \, \bm{1}_{\mathbb{R}^3} - \bm{\sigma} \cdot \nabla \bm{u} )
    = (\nabla \bm{u})^T \cdot \dot{\bm{p}}  
       - (\nabla \bm{p})^T \cdot \dot{\bm{u}}   
$$
Taking into account the identity
$$ \nabla \cdot (\dot{\bm{u}} \otimes \bm{p} ) 
    = (\nabla \cdot \dot{\bm{u}}) \, \bm{p} 
      + (\nabla \bm{p}) \cdot \dot{\bm{u}}
$$
we obtain (\ref{nabla cdot T + f =})
$$ \nabla \cdot \bm{T} 
    = (\nabla \bm{u}) \cdot \dot{\bm{p}}
      - (\nabla \cdot \dot{\bm{u}}) \, \bm{p}
      - 2 \, (\nabla_s \bm{p})  \cdot \dot{\bm{u}}
$$
where occurs the  tensor (\ref{def T}) 
$$ \bm{T} = h \, \bm{1}_{\mathbb{R}^3} - \bm{\sigma} \cdot \nabla \bm{u} -  \dot{\bm{u}} \otimes \bm{p}
$$


\end{document}